\documentclass[11pt]{article}

\usepackage{graphics}
\usepackage{amsmath, amssymb}
\usepackage{color}%
\newcommand{\degr}{$^{\circ}$}

\begin{document}

\title{Characterization of PMMA--3-octanone 
binary \\ by turbidity and light scattering measurements.} %Title of paper
%\titlerunning{Characterization of PMMA--3-octanone binary mixture}
%\authorrunning{Crauste-Thibierge et al.}

\author{Caroline Crauste-Thibierge, Cl\'emence Devailly, Audrey Steinberger, Sergio Ciliberto, \\ Laboratoire de Physique, ENS de Lyon CNRS UMR 5672\\46, all\'ee d'Italie F69007 Lyon, France}
\maketitle

\begin{abstract}{
 We measure  the coexistence  curve and the critical point properties of a binary mixture composed by polymethylmethacrylate (PMMA) at $M_w = 55900$ g/mol with 3-~octanone. This binary mixture which has a demixing transition with an upper critical solution temperature $T_c$ has indeed interesting properties which may be useful for several application : a) its correlation length is  larger than that of a liquid-liquid binary mixture, b) it is less viscous   than a polymer blend; c)  3-~octanone has an evaporation rate much smaller than other solvents.  
The mixture is first characterized by turbidity, to get the demixing temperature for different volumic fraction  of PMMA.  The coordinates of the critical point are obtained: $\phi_c = 12.8\pm0.2$ \% and $T_c = 306.5\pm0.1$ K.  The correlation length  $\xi$ near the critical point is then measured in a solution with a 12.8~\%  volumic fraction of PMMA using static light scattering. Using the fact that PMMA-octanone mixture has scaling exponents compatible with Ising 3D,  we determine more precisely the critical temperature $T_c = 306.58\pm0.04$ K and we find that  $\xi \simeq \xi_0 [(T-T_c)/{T_c}]^{-0.63}$ with $\xi_0 = 0.97\pm 0.02$~nm. The discrepancy between this value and that extrapolated from other measurements  based on  turbidity is discussed.}
\end{abstract}
%   {PACS-key}{discribing text of that key}   \and
 %     {PACS-key}{discribing text of that key}
 %    } % end of PACS codes
%

%
\section{Introduction}

There is nowadays a revival of the critical systems which are interesting  to study confinement problems \cite{Bramwell1998,Joubaud2008}, critical Casimir \cite{Hertlein2008,Gambassi2008,Dean2009}, and
out of equilibrium properties after a quench at the critical point \cite{Berthier2001,Joubaud2009b}.
Binary mixtures are quite useful and interesting for studying these problems because the demixing transition
presents a critical point which belongs to the Ising
3D universality class. Binary mixtures close to their critical point can also be used  to tune thermally and reversibly  the colloid-colloid and colloid-substrate interactions when colloids are suspended in such a mixture \cite{Soyka2008}.

For these reasons and the above mentioned applications, it is useful to have several well characterized  mixtures which allow  the tuning  of the experimental parameters.  
This study has been motivated by the fact that 
for our applications we need a mixture with a rather large correlation length, but with a low viscosity, to have quite fast demixing and remixing. Thus we choose a polymer solvent mixture: polymethylmethacrylate (PMMA) and 3-octanone. This solvent  has the advantage of having a low evaporation rate compared to other often used solvents such as cyclohexane or methylcyclohexane. This mixture has an upper critical transition temperature, being homogeneous at high temperature and demixing in a polymer rich phase and a polymer poor phase at low temperature. It has the advantage of having a critical temperature not too far from room temperature and it has already been characterized by Xia \textit{et al.}\cite{Xia1992,Xia1996,An1997} in a wide range of molecular weights going from 26 900 g/mol to 596 000 g/mol. 

The PMMA molecular mass we work with is different from the ones they measured. Thus we did turbidity measurements to check the coexistence diagram of the mixture and locate the critical concentration.  Our main interest is in measuring precisely the correlation length $\xi$ and the critical temperature $T_c$, for our applications of this mixture. Therefore we perform static light scattering measurements. We obtain $\xi = \xi_0\varepsilon^{-\nu}$ with $\varepsilon$ the reduced temperature: $\varepsilon = \frac{T-T_c}{T_c}$. In the explored temperature range the scaling behaviour corresponds to the Ising 3D model with $\nu = 0.63$. We obtain $T_c = 306.58\pm0.04$ K and $\xi_0 = 0.97\pm0.02$ nm. The $\xi_0$ value we found is different from the one extrapolated from Xia \textit{et al.} data\cite{An1997}, obtained by turbidity measurements. 

The article  is organized as follow. We describe first the experimental set up and the methods. Then we use turbidity data to get the coexistence curve of PMMA-3-octanone mixture. In section IV we characterize the growing correlation length approaching the critical point and comment on the difference between our measurements and the one from Xia \textit{et al.}. Finally we conclude in section V.

\section{Experimental set-up}
\label{SetUp}

\subsection{Materials}
3-octanone (sup. 98\%) is purchased from Sigma-Aldrich. Polymethylmethacrylate (PMMA) is also purchased from Sigma-Aldrich (Fluka, analytical standard, for GCP) with a molecular weight $M_w~=~55 900$ g$/$mol and a polydispersity $M_w/M_n~=~1.035$. This molecular weight has been chosen  because of its small polydispersity and the quite low viscosity of the mixture. It is in the molecular weight range studied by Xia \textit{et al.}~\cite{Xia1996}, therefore
we estimate $T_c = 308.9$~K %35.77$\degC{} 
and $\phi_c = 12.8$ \% for the chosen molecular weight by linear interpolation between the molecular weights 48 600 g/mol and 95 000 g/mol used by ref. \cite{Xia1996}.

We prepared different volume fractions from 6 \% to 16~\%  in polymer. The solutions are made under a laminar flow hood. In order to obtain a given volume fraction, the polymer is weighted before adding a volume of 3-octanone calculated from the density of the polymere $\rho_{PMMA}~=~1.17$ given by the supplier. The solution is then mixed at 325 K during one night to ensure a good dissolution.

\subsection{Turbidity measurements} 
\label{TurbiditeCellules}

For the turbidity measurements, the sample is a thin mixture layer between  two glass plates, 1mm thick. The thickness of the liquid layer is 150 $\mu$m, except for solutions having a polymer fraction of 16\%. Indeed, the solution becomes very viscous at high polymer fraction and increasing the thickness of the turbidity cells to 250 $\mu$m is necessary to be able to fill them completely without leaving bubbles. 
The cell's thickness is fixed by five polycarbonate spacers (see fig. 1) glued with NOA 81. Since octanone is a very reactive solvent, especially with polymers and glues, we tested different kind of glues and concluded that the more appropriate glue is UV-insulated NOA 81. So we prepare our cell with only two small openings, put Araldite on the long sides where there is no opening, and insert the mixture between the plates with a syringe. Then we close the two openings with NOA 81, and harden  it during typically 20 seconds under a very intense UV source (Opticure 4T). Finally we add Araldite on the NOA 81. NOA 81, hardened in presence of octanone, and Araldite are still slightly porous to octanone, so we can use our cells during only 3 weeks before seeing bubbles due to octanone evaporation. With pure octanone, we have to modify the cell to have only one deep opening and keep an air bubble between octanone and NOA 81, otherwise NOA 81 is dissolved in octanone before polymerizing. If the volume of the bubble is small enough, then density variations with temperature do not induce flows in the cell. This has been checked using optical microscopy.

 \begin{figure}
\resizebox{0.45\textwidth}{!}{%
\includegraphics{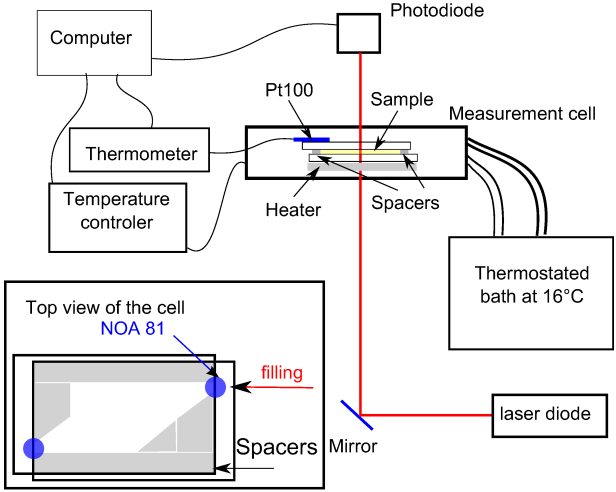}}%
 \caption{\label{SchemaTurbidite} Turbidity experiment: the laser beam goes through the sample and intensity is collected on a photodiode. In the insert, a sketch of the home-made cells we make for the turbidity experiment.}
\end{figure}

A red laser beam from a laser diode at 632.8 nm crosses the sample as shown on fig \ref{SchemaTurbidite}, and the transmitted intensity is measured with a photodiode connected to a computer via a NI USB-6009 card.  The sample is horizontal so we avoid any gravity effect. The sample's temperature is regulated by an Instec cell and temperature controller (mK1000), modified at the lab. We use a water circulation to keep the frame of the Instec cell at 289~K % 16\degC{} 
as a cold point for fast quenches. The mK1000 has a thermometer under the sample and we add a Pt100 probe glued on the top glass plate of the sample (see fig. \ref{SchemaTurbidite}), which is on the opposite side to the heating resistor. This Pt100 is calibrated thanks to an AOIP PHP601 thermometer connected to a calibrated reference Pt100 probe purchased from AOIP. We measure the temperature and the temperature gradient in our sample. The temperature stability of the cell is better than 0.02~K.
The maximum temperature difference between the bottom and the top of the sample is 0.63 K when the sample is cooled at 5 K/min, which is our faster ramp speed. At 0.01 K/min, the temperature difference between the top and the bottom is smaller than 0.15 K. 
Since the thickness of the glass plates (1mm each) is much larger than that of the sample we assume that the sample temperature is homogeneous over its thickness. 
 
We perform a series of different temperature ramps going down from 303.5 K or 311.5 K (depending on the concentration studied) to 293.5 K or 301.7 K at rates varying from 5 K/min to 0.01 K/min. Before each ramp the sample is maintained at the highest temperature during one hour which is enough to allow homogeneous mixing on our small volume. After the ramp we let the sample evolve at the lowest temperature during one or two hours before increasing temperature again.

\subsection{Static light scattering}
 \subsubsection{Measurements}

The volume fraction of PMMA in octanone used in the light scattering experiment is 12.8 \% \textit{ie} the critical concentration deduced from Xia \textit{et al.}\cite{Xia1992,Xia1996,An1997}. The PMMA-octanone solution is put in a cylindrical glass cell of inner diameter $D = 22.29\pm$ 0.02 mm. The cell is closed with a Teflon cap to avoid cap dissolution from octanone vapor. Since the cap is not perfectly airtight, the sample is replaced every week, using the same batch of solution.

We perform static light scattering using a Malvern Autosizer 4700 spectrometer\cite{TseVeKoon2012} with a green laser beam at $\lambda~=~532$~nm. We let the laser stabilize more than one hour before starting the measurements. 
The sample cell is immersed in a decaline bath ensuring both an optical index matching with the glass tube and temperature stabilization. For temperature control, the decaline bath is surrounded by a water circulation connected to a thermostated bath. The temperature stability is better than 0.02~K. The temperatures we indicate in the following are measured in the decaline bath, and calibrated with the AOIP reference Pt100 probe. At the beginning of the experiment, the bath temperature is set at $T_H \approx$~319~K ($T_H-T_c \approx $ 12 K) so that the solution remains in an homogeneous state. The temperature is then slowly decreased by steps. For each temperature step, the solution is left to equilibrate at constant temperature for 40~min before starting the data acquisition. The scattered intensity is then recorded as a function of the scattering angle $\theta$ for angles varying between 15\degr{} and 150\degr. Each spectrum is measured four times, and these four spectra are averaged. The temperature is then decreased toward the next step during a 20~min temperature ramp.

 \subsubsection{Data correction}
 \label{ScatteringCorr}

The data treatment follows the ones done by \cite{Sato1996,Gulari1972}. The scattering vector $q$ is calculated as usual: $q~=~\frac{n4\pi\sin(\theta/2)}{\lambda}$ with $n$ the refractive index of octanone ($n$~=~1.41), $\lambda$~=~532~nm the incident laser wavelength and $\theta$ the scattering angle. The acquired intensity $I_{meas}(q,T)$ is already corrected for the diffraction volume term $\sin(\theta)$ (see \cite{Chu1964}). The data should also be corrected for 1) background noise, 2) geometrical aberrations, 
3) attenuation due to the sample's turbidity, 4) dust and stray light. 

We first subtract the background intensity $I_b$ (measured with the laser off) to all our measured intensities. In order to correct the possible geometrical aberrations of the spectrometer, we measure the scattered intensity of toluene $I_{tol}(q)$ at 298 K. The structure factor of the toluene being flat on our $q$-range, the variation of the measured intensity with $q$ is due to geometrical aberrations of the spectrometer. We adjust it by a linear fit and divide all the rough intensities by the corresponding values. Then we obtain:
\begin{equation}
	I_{corr}(q,T)=\frac{I_{meas}(q,T)-I_b}{I_{tol}(q)-I_b}.
\end{equation}
 
In order to correct our data for attenuation, we cannot use the previous turbidity measurements with the red laser because turbidity depends on the wavelength of the incident light. So we replace the cylindrical cell in the light scattering setup by a square Hellma cell of side $l = 10$~mm, filled with the same sample. We measure the transmitted light power $I_t$ at $q = 0$ with a Thorlabs PM100D powermeter, following the same temperature steps as in the light scattering measurements. The measured light power varies from 50 $\mu$W at $T_H \approx$~319~K %46.00\degC{} 
to a few $\mu$W at 0.1~K above $T_c$, with a background of 0.02~$\mu$W. We normalize all the intensities to the highest temperature value $I_t(T=T_H)$ and fit them using: 
\begin{equation}
\frac{I_t(T)}{I_t(T=T_H)} = A_0\left(1-\exp\left(-\frac{T-T_0}{\Delta T}\right)\right)
\end{equation}
and get $A_0 = 0.94$, $T_0 = 306.94$ K and $\Delta T~=~0.41$~K. We use here a phenomenological expression that fits well our turbidity data especially at low temperature. We did not use eq. \ref{Tau} because we do not have the relation between the correlation length $\xi$ and the temperature at this stage of the study. In any case we checked \textit{a posteriori} that the result are not affected by the particular form chosen to fit the intensity data.  From this fit, we compute the turbidity 
\begin{equation}
\tau(T) = - \frac{\ln\left(I_t\left(T\right)/I_{t}\left(T=T_H\right)\right)}{l}
\label{TauDef}
\end{equation}
with $l$ the side of the Hellma square cell. After the background and geometrical aberrations corrections, the intensity is multiplied by $\exp\left(\tau(T)D\right)$ to compensate for attenuation within the experimental cell of diameter $D$. This correction works for single scattering but fails in case of multiple scattering, when the optical depth $\tau(T)D$ becomes higher than one\cite{Berrocal2007}. Because of the critical opalescence of our sample, which becomes highly turbid close to the critical point, this attenuation correction works only for $T-T_c~>$~0.4~K. Corrections for multiple scattering are very delicate \cite{Kao1969,Shanks1988,Lakoza1983,Schroder2003} and not done here.
Finally, we correct for the potential dust and stray light by subtracting the high temperature intensity. We obtain the intensity
\begin{equation}
\label{att_dust_corr}
I(q,T) = I_{corr}(q,T)\exp\left(\tau(T)D\right)-I_{corr}(q,T = T_H).
\end{equation}
 
\section{Coexistence curve of PMMA-octanone mixtures}

\subsection{Transition temperature}
\label{TransitionTurbidite}
In order to determine the coexistence curve of the PMMA-octanone mixture, 
we perform %different 
turbidity measurements on seven solutions of different concentrations. The ramp rates evolve from 5 K/min (which is to fast to allow temperature equilibration in the thickness of the sample) to 0.01~K/min. This is the slowest temperature rate that can be set on our temperature controller. Note that a 10 K ramp at this rate takes 17 hours.

 \begin{figure}
 \resizebox{0.45\textwidth}{!}{%
\includegraphics{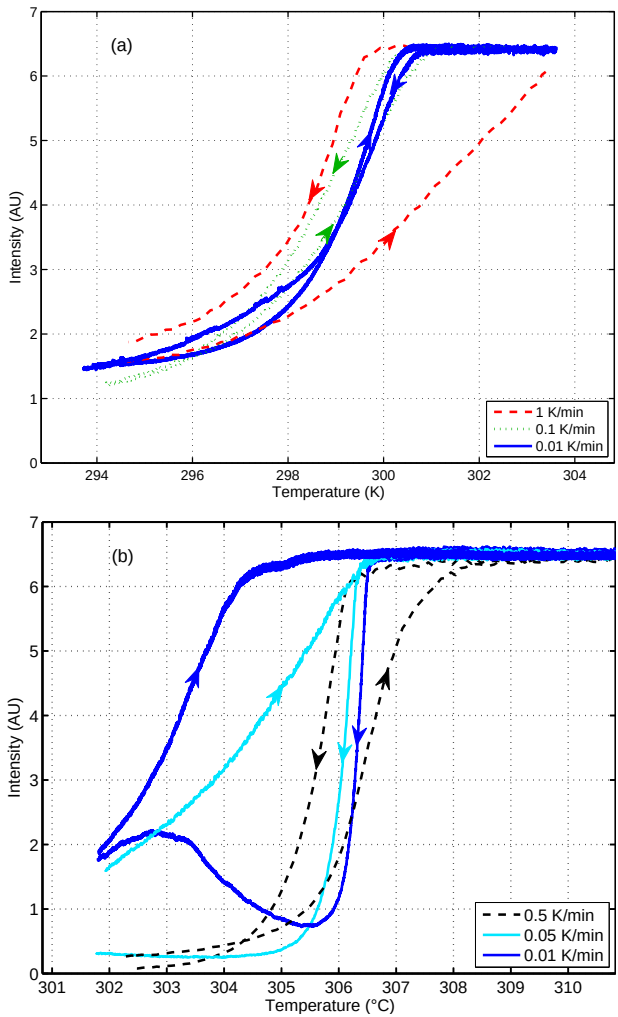}}%
 \caption{\label{Hysteresis} Transmitted light intensity during temperature cycles around the transition temperature at two different concentration: (a) 6 \% is far away from the critical point and (b) 12.6 \% is quite close.}
\end{figure}

On fig. \ref{Hysteresis}-a we show that far away from the critical point (in concentration) we have a quite large hysteresis on the transmitted intensity during a temperature cycle. Hysteresis is defined as the temperature difference between the increasing temperature ramp an the decreasing one, at half of the intensity decrease. The hysteresis decreases when the ramp rate decreases; it varies from about 2 K at 1 K/min ramp rate (at the half height of the intensity growth) to less than 0.3~K at 0.01 K/min.

\begin{figure}
\resizebox{0.45\textwidth}{!}{%
\includegraphics{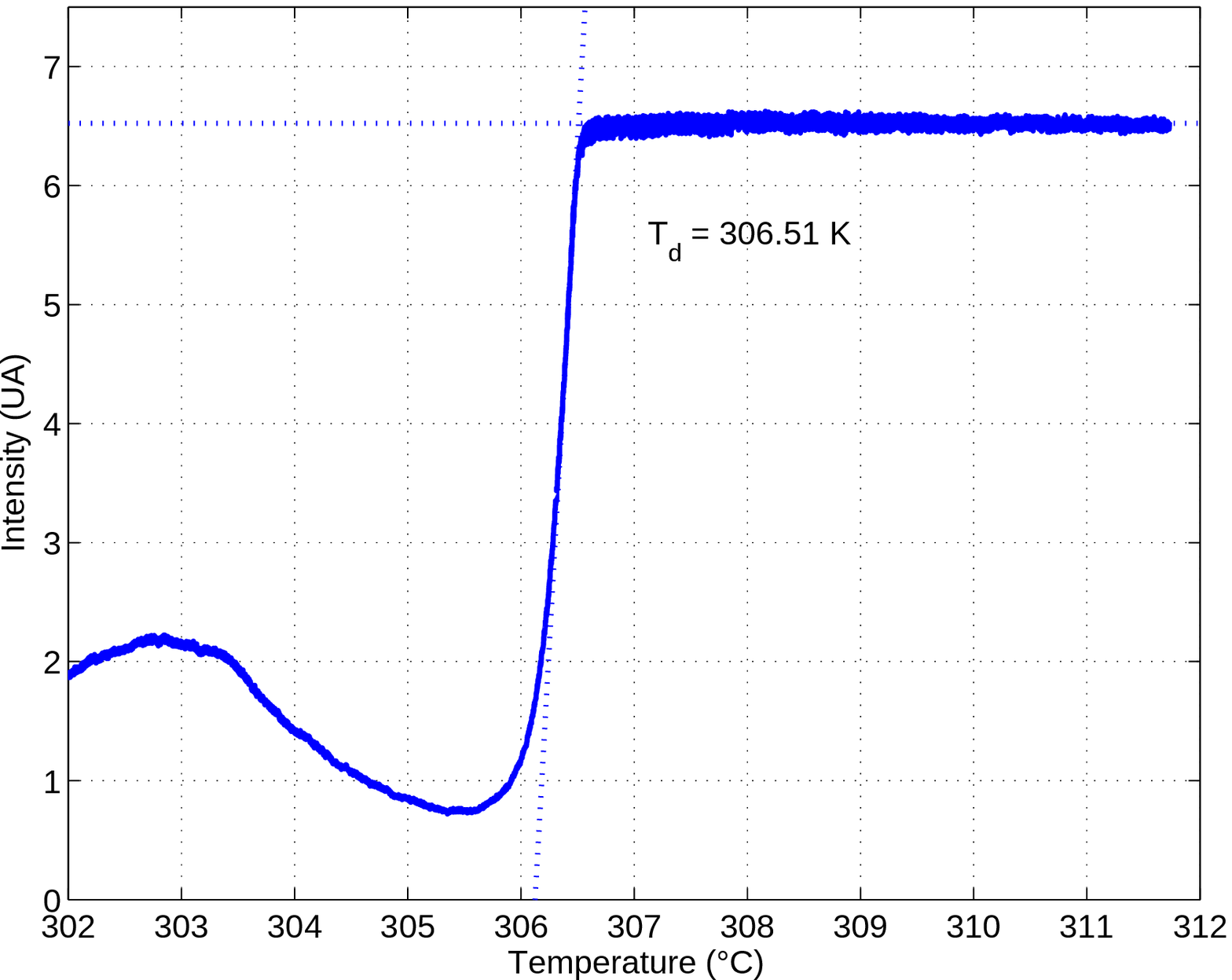}}
\\
 \caption{\label{RampDown} Intensity of the transmitted light during a temperature ramp to low temperature, at ramp rate 0.01 K/min and concentration 12.6\%. The dotted lines illustrate how $T_d$ is defined as the interception of two straight lines, as explained in the text. }
\end{figure}

Very close to the critical point, we also observe an hysteresis in the transmitted intensity during a temperature cycle, but in that case the hysteresis increases when the ramp rate decreases as shown on fig. \ref{Hysteresis}-b. 
Moreover, the evolution of the transmitted intensity can be influenced by very slow kinetic effects induced by the divergence of the relaxation time and of the correlation length at the critical point. An oscillation of the transmitted light sometimes appears at low cooling rate (see figs.~\ref{Hysteresis}-b and \ref{RampDown}). This corresponds to a phenomenon reported by D. Vollmer \textit{et al.} \cite{Vollmer2002} and explained in references\cite{Cates2003,Vollmer2008}. The main idea is that at slow cooling rate, instead of following the binodal equilibrium composition, the composition oscillates between the binodal and spinodal line. If the droplet nucleation (driven or not by gravity) is fast enough compared to the cooling rate, the medium can become more transparent before darkening again when the composition crosses the spinodal line and transits again.

We can define two transition temperatures: one when the temperature goes up $T_{up}$ and one when the temperature goes down $T_d$. At high temperature, the homogeneous initial state is always the same while the heterogeneous low temperature state depends strongly on the previous ramp rate and the waiting time at this state. As a consequence we choose to focus on the $T_d$ transition temperature.
There are several manners to define $T_d$. Tsuyomoto \textit{et al.} \cite{Tsuyumoto1984} determined the cloud point from the breakpoint of the intensity vs. time curve. But for our data, the decreasing of intensity is not sharp enough to set $T_d$ that way. Boutris \textit{et al.} \cite{Boutris1997} used the criterion of 10 \% reduction in transmittance to define $T_d$. This method is quite sensitive to noisy data and is only suitable if the decreasing slopes of the intensity vs. temperature curve for each concentration are approximately equal. As it is not the case here, we prefer to define $T_d$ as the interception of the straight line defined by the average value of the plateau at high temperature, and the line defined by a linear fit of the decreasing slope between 5 \% and 15 \% reduction in transmittance. This procedure is illustrated on fig. \ref{RampDown}. 

\subsection{Coexistence curve}
\label{DiscussionTurbidite}
 \begin{figure}
\resizebox{0.45\textwidth}{!}{%
\includegraphics{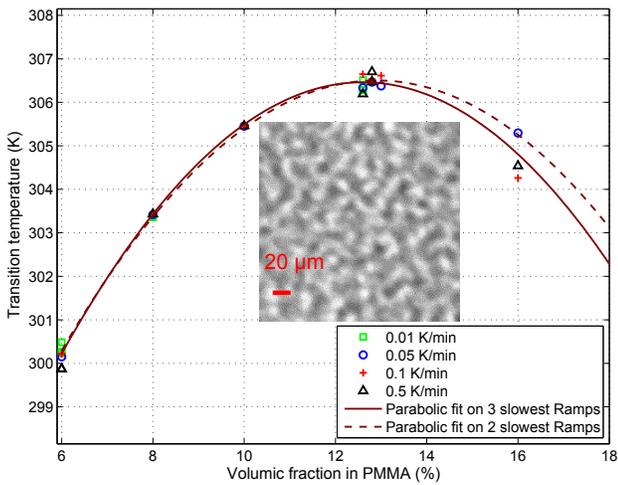}}%
 \caption{\label{CoexistenceCurve} PMMA-octanone coexistence diagram at different ramp rates with parabolic fits on the slowest ramp rates. We add a picture we took under a Leica Microscope ($\times 50$) showing the spinodal decomposition at the critical fraction, 0.2~K below the critical temperature.}
\end{figure}

From the transition temperatures $T_d$ measured at different concentrations, %represented data on fig. \ref{VitesseTrempe}, 
we built the coexistence curve of the PMMA-3-octanone binary mixture on fig \ref{CoexistenceCurve}. On this curve we cannot separate the binodal and spinodal curves. When we measure the $T_d$ transition temperature we do not exactly know the mechanism of the transition. All we can say is that close to the critical point, we have a spinodal decomposition followed by the phenomenon of oscillating demixing described in \ref{TransitionTurbidite}.
Of course, microscopy gives us a direct view of the demixing process but the technique is limited by the optical resolution of the microscope \textit{ie} a few micrometers, so that the transition is detected later than with our turbidity measurement.

Slow kinetic effects and oscillating demixing result in a relatively important dispersion of the measured transition temperatures near the critical point. Therefore, we cannot simply realize different concentrations around the critical point and choose the one which has the highest transition temperature at the slowest ramp rate. Instead, we choose to determine the critical point by using all the concentrations we measured and fitting our temperatures by a parabolic evolution in concentration. This method has the advantage to be less sensitive to the way the system takes through demixing and to take into account all the data we have collected. This implies that the coexistence curve is symmetric around the critical point, which has already been shown by Xia \textit{et al.} \cite{Xia1996}. We fitted the data corresponding to the two slowest ramps (0.01~K/min and 0.05~K/min) or to the three slowest ramps (adding 0.1 K/min) and get the values: $\phi_c =~12.9$~\%, $T_c =  306.5\pm0.1$~K and  $\phi_c = $ 12.6~\%, $T_c =  306.5\pm 0.1$~K respectively. 
These values of the critical concentration $\phi_c$ are compatible with the value $\phi_c =~12.8$ \% infered from the work of Xia \textit{et al.}~\cite{Xia1996}, which is the reason why we choose to use the latter concentration for the light scattering measurements. 

Our critical temperature is 2.42 K smaller than that in ref\cite{Xia1996}. This may come from its great sensibility to the smallest amount of impurities in the mixture. We did not use the same PMMA and octanone as Xia \textit{et al.}, so maybe our sample contains different impurities with a different effect on $T_c$. Note that, we carefully calibrated our temperature probe with a reference probe.

\section{Growing correlation length approaching the critical point}

\subsection{Static light scattering}
\label{GrowingXi}
\subsubsection{Ornstein-Zernike model}

 \begin{figure}
\resizebox{0.45\textwidth}{!}{%
\includegraphics{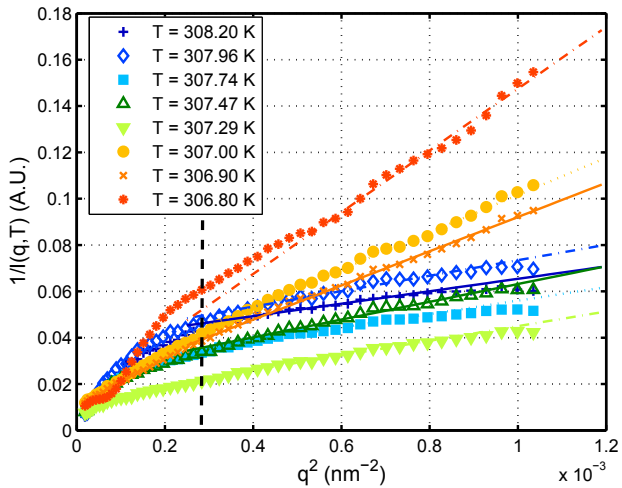}}%
 \caption{\label{OZdiag} Ornstein-Zernike plot of the inverse of the scattered light intensity as a function of $q^2$. All the corrections are done except for multiple scattering. The points are our experimental data, the lines are the fits over the $q$-range at the right of the dashed line.}
\end{figure}

It is very well known that binary mixtures belong to the Ising 3D universality class. Near the critical point, the scattered light intensity depends on the long range correlation effects of the density (or concentration) fluctuations. In the single scattering limit, 
the Ornstein, Zernike and Fisher model \cite{Ornstein1914,Debye1960,Debye1963,Fisher1967} 
expresses the scattered light intensity as:

\begin{equation}
\label{OZF_eq}
I(q,T) = \frac{I_0(T)}{\left(1+q^2\xi(T)^2\right)^{1-\eta/2}},
\end{equation}
with $I_0 = A_1\chi_T$. The factor $A_1$ is insensitive to temperature but also contains the possible fluctuations of the laser intensity. $\chi_T$ is the isothermal osmotic compressibility scaling like 
\begin{equation}
\chi_T = \chi_{T,0}\varepsilon^{-\gamma}
\label{ScalingI0}
\end{equation} 
where $\gamma$ is the critical exponent for $\chi_T$ and 
\begin{equation}
\varepsilon = \frac{T-T_c}{T_c}.
\end{equation}
$\xi$ is the static correlation length scaling like
\begin{equation}
\xi = \xi_0\varepsilon^{-\nu}
\label{ScalingXi}
\end{equation}
and $\eta$ is a small correction factor, also given by the hyperscaling relation
\begin{equation}
\gamma = (2-\eta)\nu.
\end{equation}
The values of the Ising model exponents $\gamma$, $\nu$ and $\eta$ are given in table \ref{TableExponents}, this comparison is done only to check the accuracy of our fits.

\begin{table}
%\resizebox{0.45\textwidth}{!}{%
\begin{tabular}{|c|c|c|}
\hline\noalign{\smallskip}
Exponent & Ising universality class & PMMA/octanone mixture \\
\noalign{\smallskip}\hline\noalign{\smallskip}
$\nu$ & 0.63 & 0.61 $\pm0.03$ \\
\noalign{\smallskip}\hline\noalign{\smallskip}
$\gamma$ & 1.24 & 1.25$\pm$ 0.20\\
\noalign{\smallskip}\hline\noalign{\smallskip}
$\eta$ & 0.03 & --\\
\noalign{\smallskip}\hline
\end{tabular}
\caption{\label{TableExponents} Values of the critical exponents of the Ising model in three dimensions compared to the exponents measured in the PMMA/octanone mixture.}
\end{table}

In a first approximation we neglect $\eta$ in order to determine $I_0(T)$ and $\xi(T)$ in an Ornstein-Zernike framework \cite{Ornstein1914}, where $I^{-1}$ is modeled as an affine function of $q^2$:
\begin{equation}
I(q,T)^{-1} =I_0(T)^{-1}\left[1+q^2\xi(T)^2\right] 
\label{EqOZ}
\end{equation}

The fig. \ref{OZdiag} shows the evolution of $I^{-1}$ with $q^2$ for temperatures varying from 308.2 K to 306.8 K.
It is almost linear, following the Ornstein-Zernike behaviour, except in the low $q$-range where multiple scattering becomes important. The $q$-range affected by multiple scattering increases when the temperature gets closer to the critical temperature $T_c$. Using eq.~\ref{EqOZ}, a linear fit of the experimental data in the high $q$-range (defined on fig.~\ref{OZdiag}) yields the values of $I_0$ and $\xi$ for each temperature.

 \begin{figure}
\resizebox{0.45\textwidth}{!}{%
\includegraphics{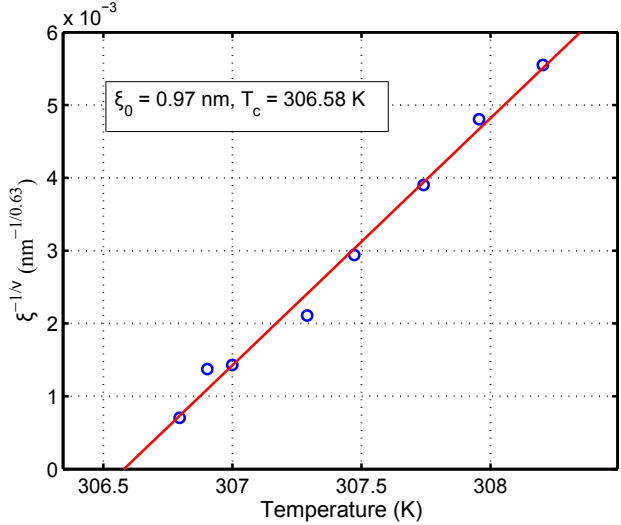}}%
 \caption{\label{KsiT} Temperature dependence of $\xi^{-1/\nu}$, with $\nu = 0.63$.  The linear fitting of the data gives $\xi_0 = 0.97\pm0.02$ nm and $T_c = 306.58\pm0.04$ K. 
}
\end{figure}

\subsubsection{Results}
\begin{figure}
\resizebox{0.45\textwidth}{!}{%
\includegraphics{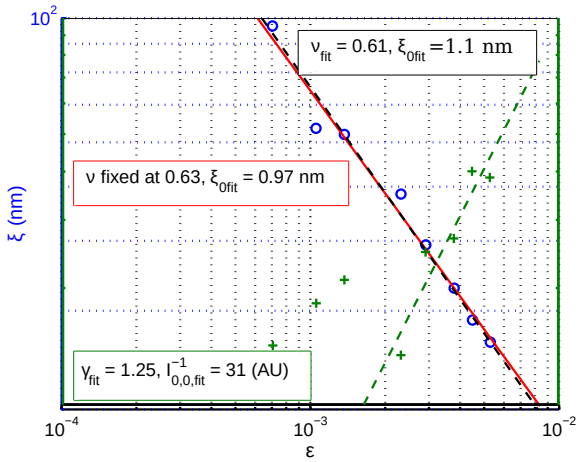}}%
 \caption{\label{LogLogXiI0} Double logarithmic plot of $\xi$ (circles, left axis) and $I_0^{-1}$ (crosses, right axis)  as a function of the reduced temperature $\varepsilon = \frac{T-T_c}{T_c}$. Letting the $\nu$ exponent free while fitting $\xi = \xi_0\varepsilon^{-\nu}$ gives $\nu = 0.61\pm0.03$ and $\xi_0 = 1.1\pm0.2$  nm (black dashed line). Fixing $\nu = 0.63$ gives $\xi_0 = 0.97\pm 0.02$ nm (plain line). Fitting $I_0 = I_{0,0}\varepsilon^{-\gamma}$ gives $\gamma = 1.25\pm0.20$ and $I_{0,0}^{-1} = 31\pm2$ AU, when it is fitted on the range where the optical depth $\tau D$ is smaller than 1, \textit{ie.} $\varepsilon > 1.4.10^{-3}$.   %
 }
\end{figure}

We determine the critical temperature $T_c$ from the evolution of $\xi^{-1/\nu}$ as a function of temperature given by eq. \ref{ScalingXi}. When temperature goes down to $T_c$, the correlation length $\xi$ diverges, so $\xi^{-1/\nu}$ vanishes at the critical temperature (see fig. \ref{KsiT}). Assuming $\nu = 0.63$, we get $\xi_0 = 0.97\pm0.02$~nm and $T_c = 306.58\pm0.04$K. Fig.~\ref{LogLogXiI0} shows a double logarithmic plot of $\xi$ as a function of the reduced temperature $\varepsilon = \frac{T-T_c}{T_c}$ with $T_c = 306.58$~K, in order to further check the scaling given in eq. \ref{ScalingXi}. If we let the value of $\nu$ as a free fit parameter, we get $\nu = 0.61\pm0.03$ and $\xi_0 = 1.1\pm0.2$~nm, which is compatible with the value obtained by fixing $\nu = 0.63$. 

The intensity $I_0$ is supposed to scale like eq.~\ref{ScalingI0} but the values we have are too dispersed to calculate $T_c$ from $I_0$ with a reasonable accuracy. We plot on the right axis of fig.~\ref{LogLogXiI0} $I_0^{-1}$ as a function of the reduced temperature $\varepsilon = \frac{T-T_c}{T_c}$ deduced from the correlation length measurements. We can distinguish two regions. For $\varepsilon > 1.4\cdot10^{-3}$, $I_0^{-1}$ is compatible with the scaling behaviour predicted by eq. \ref{ScalingI0} with an exponent $\gamma = 1.24$; a fit of $I_0^{-1}(\varepsilon)$ in this region with the exponent $\gamma$ as a free parameter yields $\gamma = 1.25\pm0.20$. The dispersion of the data is due to slow fluctuations of the laser intensity with a standard deviation of 7\%.  
For $\varepsilon < 1.4\cdot10^{-3}$, the data depart from this scaling behaviour. In this region close to the critical point, the optical depth $\tau D$ is higher than 1 due to critical opalescence, and the attenuation correction fails as explained in section \ref{ScatteringCorr}. Since the dust and stray light correction by $I_{corr}(q,T = T_H)$ in eq.~\ref{att_dust_corr} is small in this region, a bad attenuation correction mainly results in a wrong multiplicative coefficient to the intensity $I(q,T)$. This affects the measurement of the intensity $I_0$ by our fitting procedure, but not the one of the correlation length $\xi$. This is the reason why the measured $\xi$ follow the expected scaling law even in the $\varepsilon < 1.4\cdot10^{-3}$ region.

\begin{figure}
\resizebox{0.45\textwidth}{!}{%
\includegraphics{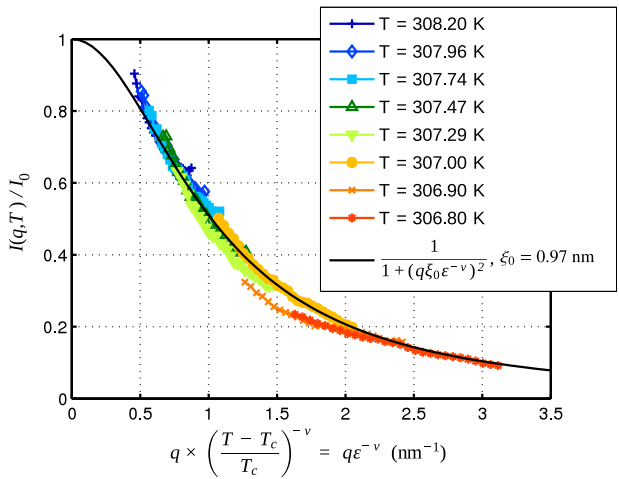}}%
 \caption{\label{Collapse} Collapse of the scattered intensities: $I(q,T)/I_0(T)$ is plotted as a function of $q\varepsilon^{-\nu}$, confirming the critical behaviour of the transition. Data are given over the $q$-range we used for the Ornstein-Zernike fit.}
\end{figure}

Finally, we plot the reduced intensity $I(q,T)/I_0(T)$ as a function of $q\varepsilon^{-\nu}$ on fig.~\ref{Collapse} in order to check that the data collapse on a master curve, as can be expected from relationships \ref{OZF_eq} and \ref{ScalingXi}. The data nicely collapse over all the $T$ and $q$ ranges confirming that our sample has a critical behaviour over all this range.

\subsection{Correlation length from turbidity}

Xia \textit{et al.} \cite{An1997} already measured the correlation length in a series of PMMA-octanone mixtures with different PMMA molecular weights. They measured $\xi_{0,\tau}$ from turbidity data at $\lambda = $ 632.8 nm and get $\xi_{0,\tau} = 0.5778 \pm 0.013$~nm at $M_w =$~48600~g/mol and $\xi_{0,\tau} = 0.6255 \pm 0.017$~nm at $M_w =$~95000~g/mol while we measured $\xi_{0} = 0.97  \pm 0.02$~nm at $M_w = $ 55 900 g/mol.  

Turbidity measurements have been used to determine the correlation length of several different mixtures\cite{Chen2001,Kita1997,An1998,Yin2012}. Turbidity is defined as the attenuation of transmitted light per unit optical path length (see eq. \ref{TauDef}). When the sample does not absorb light, turbidity results from scattering.
In the single scattering regime, turbidity is linked to $\xi_{\tau}$ by an integrated form of the Ornstein-Zernike eq. \cite{Puglielli1970,Kita1997,Chen2001}
\begin{equation}
\tau = A_2 \chi_T G(z) = A_2\chi_{T,0}\varepsilon^{-\gamma}G(z)
\label{Tau}
\end{equation}
with $A_2$ treated as a constant with temperature and 
\begin{equation}
G(z) = \displaystyle{\frac{2z^2+2z+1}{z^3}\ln(1+2z)-\frac{2(1+z)}{z^2}}
\end{equation}
with $z =2\displaystyle{\left(\frac{2\pi n \xi}{\lambda}\right)^2}$. 

Getting $\xi_0$ supposes to adjust eq. \ref{Tau} with five parameters: $A_2\chi_{T,0}$, $\xi_0$, $T_c$, $\gamma$ and $\nu$. Even if the $\gamma$ and $\nu$ values are fixed, the fit is very sensitive to small variations of the other parameters and especially of the value of $T_c$. This is why we choose to measure $\xi_0$ \textit{via} static light scattering measurements. 

Sato \textit{et al.}\cite{Sato1996} determined $\xi_0$ both  by turbidity and static light scattering on a PDMS-PEMS mixture, which has a low turbidity due to very close refractive indices of these polymers. The value obtained by turbidity measurements is $\xi_{0,\tau} = 1.37 \pm 0.35$ nm while the value obtained by static light scattering is $\xi_{0} = 1.62 \pm 0.05$ nm. This 25~\% difference between the two results is compatible with the difference between the $\xi_0$ we measured by static light scattering and the one  from Xia \textit{et al.}. 

Following Sato \textit{et al.}, we believe that the static light scattering is a more precise method to measure correlation lengths than turbidity. On the one hand it allows an independent determination of $T_c$, on the other hand it allows to better deal with multiple scattering in a highly turbid mixture like  PMMA-octanone.  Near the critical point the small angle signal is greatly affected by multiple scattering. This implies that, with static light scattering, the $q$-range used for the Ornstein-Zernike fit becomes smaller, but it does not affect the precision too much.  With turbidity measurements, the expression used to analyze the results is only valid in the single scattering approximation, so one has to be very careful as soon as multiple scattering affects small angles.%Two correlation lengths can only be compared if they have been measured with the same technique.
 
\subsection{Specific properties and universality class of PMMA-octanone mixture}

\begin{table}
%\resizebox{0.45\textwidth}{!}{%
\begin{minipage}{0.45 \textwidth}
\begin{tabular}{|c|c|c|c|}
\hline\noalign{\smallskip}
Mixture & $\xi_0$ (nm) & $M_w$ (polymers) & ref \\
 & & (g/mol) & \\ 
\noalign{\smallskip}\hline\noalign{\smallskip}
SF$_6$ & $0.15\pm0.023$& -- & t\cite{Puglielli1970} \\
\noalign{\smallskip}\hline\noalign{\smallskip}
Lutidine & $0.3\pm0.02$ & - & s\cite{Gulari1972} \\
water & & & \\
\noalign{\smallskip}\hline\noalign{\smallskip}
triethylamine & $\approx1$ & - & \cite{Lefort2011} \\
water & & & \\
\noalign{\smallskip}\hline\noalign{\smallskip}
polystyrene & 0.4822$\pm$ 0.0026 &10100 &t\cite{Zhou2002} \\
methylcyclohexane & & & \\
\noalign{\smallskip}\hline\noalign{\smallskip}
polystyrene & 0.7871$\pm$ 0.0080 &330020 &t\cite{Zhou2002} \\
methylcyclohexane & & & \\
\noalign{\smallskip}\hline\noalign{\smallskip}
\textit{d}-polybutadiene & 0.70 & 232000 & \cite{Schwahn1987,Janssen1992}\\
polystyrene & & 89000 & \\
\noalign{\smallskip}\hline\noalign{\smallskip}
PDMS & 1.62$\pm$0.05 &$1.91.10^4$& s\cite{Sato1996}\\
PEMS & & $1.40.10^4$  &  \\
\noalign{\smallskip}\hline\noalign{\smallskip}
PDMS & 1.37$\pm$0.35 &$1.91.10^4$& t\cite{Sato1996}\\
PEMS & & $1.40.10^4$  &  \\
\noalign{\smallskip}\hline
\end{tabular}
\caption{\label{TableXi}Values of $\xi_0$ for different critical systems from literature. The letter s or t in the last column indicates if the correlation length is measured from turbidity (t) or static light scattering (s), or another technique if there is no letter. $M_w$ is the molecular mass of polymers.}
\end{minipage}
\end{table}

 In order to check the quality of our data we checked that the binary mixture of PMMA-octanone belongs to the universality class of Ising 3D like many other polymer binary mixtures or polymer-solvent mixtures. In the temperature range we study, the sample always displays critical behaviour, as we do not see any change in the slopes of fig. \ref{LogLogXiI0}. We do not see the cross-over between critical and mean-field behaviour. This is not surprising since Herkt-Maetzky and Schelten \cite{Herkt1983} only see it  in the region where $\varepsilon > 2.10^{-2}$ in a deutero-polystyrene and polyvinylmethylether mixture. In a PDMS/PEMS mixture, Sato \textit{et al.}  did not see any cross-over for $\varepsilon < 2.10^{-2}$. Our values are in the $\varepsilon \leq 5.10^{-2}$ range.

This study was motivated by the need of finding a binary mixture with a large $\xi_0$ but not very viscous. This last criteria excludes polymer mixtures. Increasing the molecular weight of polymers in  mixtures does not increase $\xi_0$ very much. In a binary mixture of polymers where one molecular weight is kept constant and the other one is varying, $\xi_0$ scales like\cite{An1998CPL}:
\begin{equation}
\xi_0 \propto  M_w^n,
\end{equation}
with $M_w$ the molecular mass of the polymer and $n~=~0.18$. The behaviour is the same for a binary mixture of a polymer and a solvent: Xia \textit{et al.} found $n = 0.15 \pm0.02$ \cite{An1997}. Taking very heavy polymers increases mostly the viscosity, faster than the correlation length. This study shows that PMMA-3-octanone has quite a high $\xi_0$ value (see table \ref{TableXi}) but is still easy to manipulate.

\section{Conclusion}
In this paper, we investigate the coexistence curve and static critical behaviour of a polymer solvent mixture of PMMA-3-octanone. We obtain the coexistence curve of the mixture from turbidity measurements performed at very slow ramp rate (0.01 K/min). The slow ramp rate avoids temperature gradients in the sample and allows us to be as close to equilibrium as possible. Since the decrease of transmitted light intensity is not very sharp at the transition, we choose a phenomenological way of determining the transition temperature on decreasing temperature ramps. The critical volumic fraction is compatible with the value of $\phi_c = 12.8\pm 0.2$ \% infered from the results from Xia \textit{et al.}\cite{Xia1996}, and  the critical temperature is 306.58$\pm$0.05 K from our calibrated reference temperature probe.

We used a $\phi_c = 12.8 $ \% sample to perform static light scattering and get $T_c = 306.5\pm0.1$ K and $\xi_0~=~0.97\pm0.02$~nm with $\nu =0.63$. The value of $T_c$ obtained by static light scattering is more precise than the one obtained from the coexistence curve. However since both are compatible, this validates \textit{a posteriori} our determination of  the transition temperature from the turbidity measurements.

The values of the exponents $\nu$ and $\gamma$ are compatible with the Ising 3D model and we did not find any indication of the crossover between the critical behaviour and the mean-field behaviour, however this is not surprising considering our $\varepsilon$ range. 

The $\xi_0$ values we found is 30 \% higher than the value  from Xia \textit{et al.}, obtained by turdidity measurements. Sato  \textit{et al.} found the same difference on a PDMS-PEMS mixture \cite{Sato1996}, so we think that this results from a bias of the chosen method.
We believe static light scattering is a more reliable and precise method to get a correlation length because it is based on large $q$-range measurements, that are less affected by multiple light scattering than $q$ = 0 measurements. It also has the big advantage to allow simultaneously an independent measure of $T_c$ and $\xi_0$.

If we compare the $\xi_0$ value of PMMA-octanone to different other mixtures, we find that it is an interesting mixture having a high $\xi_0$ with a quite small molecular mass of the polymer and a low viscosity solvent. This makes PMMA-octanone a very versatile mixture for future applications.
\\
\\
\\
\textbf{Acknowledgments} We are grateful to  Artyom Petrosyan, Eric Freyssingeas and Patrick Oswald for their precious technical help, and to Hugo Jacquin and David Lopes Cardozo for the fruitful discussions we had.  This research was supported by the European Research Council Grant OUTEFLUCOP.

%
% BibTeX users please use
% \bibliographystyle{}
% \bibliography{}

%Author, \textit{Book title} (Publisher, place year) page numbers
%\end{thebibliography}

\end{document}